\documentclass{article}
\usepackage[T1]{fontenc}
\usepackage{url,color,amstext,amsmath,amssymb,amsthm,tikz,tikz-inet,pgf,pgfplots}
\usepackage{fp,float}
\usepackage[capitalise]{cleveref}
\usepackage[super]{nth}
\usepackage{a4wide}

\def\colorpat{red}
\def\colorbruno{blue} 

\newcommand\word[1]{\texttt{#1}}

\newif\ifnotes\notestrue
\def\boxnote#1#2{\ifnotes\fbox{\footnote{\ }}\ \footnotetext{ From #1: #2}\fi}
\newcommand\mpat[1]{{\color{\colorpat} #1}}
\newcommand\spat[1]{{\scriptsize{\color{violet} #1}}}
\newcommand\hspat[1]{}
\def\pat#1{\boxnote{Patrick}{\color{\colorpat}#1}}
\def\hpat#1{}
\newcommand\mbruno[1]{{\color{\colorbruno} #1}}
\newcommand\mbrunosmall[1]{{\scriptsize{\color{\colorbruno} #1}}}
\def\bruno#1{\boxnote{Bruno}{\color{\colorbruno}#1}}
\def\hbruno#1{}

\newcommand\ind[1]{1\hspace{-.55ex}\mbox{\textnormal l}_{\{#1\}}}  

\gdef\figwidth{7cm}
\gdef\figheight{6cm}
\gdef\bc{b^{\rm c}}
\gdef\bv{b^{\rm v}}
\gdef\subc{{\tilde b}^{\rm c}}
\gdef\subv{{\tilde b}^{\rm v}}
\gdef\suv{{\tilde v}}
\gdef\pc{p^{\rm c}}
\gdef\pv{p^{\rm v}}
\gdef\CTR{\text{\footnotesize CTR}}
\gdef\dd{\mathrm{d}}
\gdef\define{:=}
\gdef\Ealph{\bar{\alpha}}
\gdef\Ebet{\bar{\beta}}
\gdef\Ealphbet{\overline{\alpha\beta}}

\renewcommand\S[1]{S_{\mathrm{#1}}}
\newcommand\Rev[1]{R_{\mathrm{#1}}}

\def\E {{\mathbb E}}
\def\EE {{\mathbb E}}
\def\FF {{\mathbb F}}
\def\II {{\mathbb I}}
\def\LL {{\mathbb L}}
\def\NN {{\mathbb N}}
\def\PP {{\mathbb P}}
\def\RR {{\mathbb R}}
\def\ZZ {{\mathbb Z}}

\pgfcreateplotcyclelist{\mylist}{
	{blue,thick},
	{red,thick}, 
	{violet,thick}, 
	{blue,thick,densely dashed},
	{red,thick,dashed},
	{violet,thick,dashed},
	{blue,thick,densely dotted},
	{red,thick,dotted}, 
	{blue,thick,dash pattern=on 6pt off 1pt on 2pt off 1pt},
	{red,thick,dash pattern=on 6pt off 1pt on 2pt off 1pt},
	{violet,thick,dash pattern=on 6pt off 1pt on 2pt off 1pt},
}

\title{%
Evaluating search engines and defining a consensus implementation
}

\author{Ahmed Kamoun\\ IMT Atlantique \\ Rennes, France \\
       \tt{ahmed.kamoun@imt-atlantique.net}
\and Patrick Maill\'e\\ IMT Atlantique \\ Rennes, France \\
       \tt{patrick.maille@imt.fr}
\and Bruno Tuffin \\ Inria Rennes Bretagne Atlantique \\ Campus Universitaire de Beaulieu, 35042 Rennes Cedex, France \\
       \tt{bruno.tuffin@inria.fr}
       }
\begin{document}
\maketitle

\begin{abstract}
Different search engines provide different outputs for the same keyword. This may be due to different definitions of relevance, and/or to different knowledge/anticipation of users' preferences, but rankings are also suspected to be biased towards own content, which may prejudicial to other content providers. In this paper, we make some initial steps toward a rigorous comparison and analysis of search engines, by proposing a definition for a consensual relevance of a page with respect to a keyword, from a set of search engines. More specifically, we look at the results of several search engines for a sample of keywords, and define for each keyword the visibility of a page based on its ranking over all search engines. This allows to define a score of the search engine for a keyword, and then its average score over all keywords. Based on the pages visibility, we can also define the consensus search engine as the one showing the most visible results for each keyword. 
We have implemented this model and present an analysis of the results.
\end{abstract}

\section{Introduction}

 Search Engines (SEs) play a crucial rule in the current Internet world. If you wish to reach some content, except if you have a specific target in mind, you dial keywords on an SE through a web browser to discover the (expected) most relevant content. The number of searches worldwide per year is not precisely known, but just talking about Google, it is thought that they handle at least two trillions of requests per year, and that it can even be much more than that\footnote{\url{https://searchengineland.com/google-now-handles-2-999-trillion-searches-per-year-250247}}. As a consequence, if you are a small content provider or a new comer, your visibility and business success will highly depend on your ranking on SEs. 
 
SEs are regularly accused of biasing their rankings\footnote{See for example \\ {\scriptsize \url{https://www.technologyreview.com/s/610275/meet-the-woman-who-searches-out-search-engines-bias-against-women-and-minorities/}}} by not only trying to provide as an output an ordered list of links based on \emph{relevance}, but to also include criteria based on revenues it could drive. The problem was brought in 2009 by Adam Raff, co-founder of the price-comparison company Foundem, saying that Google was voluntarily penalizing his company in rankings with respect to Google's own services. Such a behavior would indeed be rational from the SE perspective, as it could yield significant revenue increases; a mathematical model highlighting the optimal non-neutral--i.e., not based on relevance only--strategies of SEs is for example described in \cite{LMST17}.
The issue led to the use of expression \emph{search neutrality debate}, in relation to the \emph{net neutrality debate}  where Internet Service Providers are accused of differentiating service at the packet level to favor some applications, content, or users. Indeed, similarly, new valid content can hardly be reached if not properly considered by SEs. This is now an important debate worldwide \cite{Tuffin-ARCEP,maille2014telecommunication,wright2012defining}. 
But while defining a neutral behavior of ISPs at the network level is quite easy, a neutral behavior for SEs involves having a clear definition of relevance. Up to now this relevance is defined by SE-specific algorithms such as PageRank \cite{ilprints422}, that can additionally be (and are claimed to be) refined by taking into account location, cookies, etc. The exact used algorithms and their relation to relevance are sometimes hard to know without requiring a total transparency of SEs and free access to their algorithms, which they are reluctant to disclose.

Because of the different used algorithms, it is interesting to compare search engines, for example by giving them a grade (or score). It is often said that if someone is not happy, she can just switch, she is just one click away from another SE. But while it could be true in a fully competitive market, it is not so easy in practice with SEs since most people just know one or two SEs and do not have a sufficient expertise to evaluate them and  switch. As of May 2018, Statcounter Global Stats\footnote{\url{http://gs.statcounter.com/search-engine-market-share}} gives worldwide a market share of 
90.14\% to Google, 3.24\% to Bing, 2.18\% to Baidu, 2.08\% to Yahoo!...\hpat{Pas sur de suivre : c'est a cause des parts de marche inegales qu'on n'est pas vraiment "one click away"?}%

Our paper has several goals:
\begin{itemize}
\item First, to propose a so-called \emph{consensus SE}, defined such as some ``average'' behavior of SEs, based on the idea that this average SE should be closer to one truly based on relevance. Considering a list of several SEs, we give a score to all the provided links, by weighing them by the position click-through-rate on each SE, estimating the position-dependent probability to be clicked. The consensus SE then ranks links according to their score.
\item To give a score to SEs, comparing their results to the consensus SE.   From the score of links, we can give a score to SEs by summing the scores of the presented lists weighted by their positions. It then allows us to rank the SEs themselves and show which one seems the closest to the ``optimal'' consensus SE, for a single keyword and for a list of keywords.  
\item To discuss and compare the behavior of SEs with respect to requests in practice. We have implemented and tested our model, computing grades for SEs and distributions of scores in terms of requests. From the rankings of SEs for any keyword, we can also investigate if  there is a suspect deviation of some SEs toward their own content with respect to competitors. This would help to detect violations to a (potential) search neutrality principle.  
\end{itemize} 
Note that algorithms comparing rankings exist in the literature, see~\cite{MOWSHOWITZ20051193} and  references therein, based on differences between vectors, but there is to our knowledge no algorithm like ours taking into account the click-through-rates (weights) associated to positions.

The rest of the paper is organized as follows. 
\cref{sec:model} presents the model: the definition of link scores for any keyword, the corresponding SE score as well as the consensus SE.
\cref{sec:numer} presents an implementation of this model in practice and compares the most notable SEs.
Finally, \cref{sec:conc} concludes this preliminary work and introduces the perspectives of extension.

\section{Scoring model and consensus search engine} \label{sec:model}

We consider $n$ SEs, $m$ keywords representative of real searches, and a finite set of $\ell$ pages/links corresponding to all the results displayed for the whole set of searches.
When dialing a keyword, SEs rank links. We will limit ourselves to the first displayed page of each SE, considered here for simplicity the same number $a$ for all SEs, but we could consider different values for each SE, and even $a=\ell$.

The ranking is made  according to a score assigned to each page for the considered keyword. This score is supposed to correspond the relevance of the page. 
According to their rank, pages are more or less likely to be seen and clicked. The probability to be clicked is called the click-through-rate (CTR) \cite{maille2012overview}; it is in general SE-, position- and link- dependent, but we assume here for convenience, and as commonly adopted in the literature, a separability property:   the CTR of link $i$ at position~$l$ is the product $q'_i q_l$ of two factors,  $q'_i$ depending on the link~$i$ only, and  $q_l$ depending on the position~$l$ only. 
We typically have $q_1\geq q_2\geq \cdots \geq q_a.$
The difficulty is that the link-relative term $q'_{i}$, upon which a ``neutral'' ranking would be based, is unknown. But the position-relative terms $q_l$ can be estimated, and we assume in this paper that they are known and the same on all SEs, i.e, that SEs' presentation does not influence the CTR. We then make the following reasoning:
\begin{itemize}
\item SEs favor the most relevant (according to them) links by ranking them high, that is, providing them with a good position and hence a high \emph{visibility}, which can be quantified by the position-relative term $q_l$;
\item for a given keyword, a link that is given a high visibility by all SEs is likely to be ``objectively relevant''. Hence we use the average visibility offered by the set of SEs to a link, as an indication of the link relevance for that keyword. We will call that value the \emph{score} of the link for the keyword, which includes the link-relative term $q'_{i}$.
\item We expect that considering several SEs will average out the possible biases introduced by individual SEs, when estimating relevance; also, the analysis may highlight some SEs that significantly differ from the consensus for some sensitive keywords, which would help us detect non-neutral behaviors.
\end{itemize}

\subsection{Page score}

The notion of score of the page as defined by SEs is (or should) be related to the notion of relevance for any keyword. As briefly explained in the introduction, the idea of relevance is subjective and depends on so many possible parameters that it can hardly be argued that SEs do not consider a valid definition without knowing the algorithm they use. But transparency is very unlikely because the algorithm is the key element of their business.

In this paper, as explained above we use a different and original option for defining the score, as the 
exposition (or visibility) provided by all SEs, which can be easily computed.

Formally, for page $i$ and keyword $k$, the score is the average visibility over all considered SEs:
\begin{equation}\label{eq:def_page_score}
R_{i,k}\define \frac{1}{n} \sum_{j=1}^n q_{\pi_j(i,k)}
\end{equation}
where $\pi_j(i,k)$ denotes the position of page~$i$ on SE~$j$ for keyword $k$.
In this definition, if a page is not displayed by an SE, the CTR is taken as 0. Another way to say it is to define a position $a+1$ for non displayed pages, with $q_{a+1}=0$.

\hpat{By averaging the score links over all SEs, it is expected that bias is limited.}

\subsection{Search engine score}

Using the score of pages (corresponding to their importance), we can define the score of an SE~$j$ for a given keyword~$k$
as the total ``page score visibility'' of its results for that keyword. Mathematically, that SE score $S_{j,k}$ can be expressed as
\[
S_{j,k}\define \sum_{\text{pages }i}q_{\pi_j(i,k)}R_{i,k},
\]
where again $q_p=0$ if a page is ranked at position $p\geq a+1$ (i.e., not shown), and for each page $i$, $R_{i,k}$ is computed as in \cref{eq:def_page_score}.

\gdef\pii{\tilde\pi}
The SE score can also be computed more simply, by just summing on the displayed pages:
\begin{equation}
S_{j,k} = \sum_{p=1}^a q_{p} R_{\pii_j(p,k),k}
\end{equation}
where $\pii_j(p,k)$ is the page ranked at the $p^{\text{th}}$ position by SE $j$ for keyword $k$, i.e., $\pii_j(\cdot,k)$ is the inverse permutation of $\pi_j(\cdot,k)$. 

The higher an  SE ranks highly exposed pages, the higher its score. The score therefore corresponds to the exposition of pages that are well-exposed on average by SEs.

To define the score of SE~$j$, for the whole set of keywords, we average over all keywords:
\begin{equation}\label{eqn:SEscore}
S_{j}\define\frac{1}{m}\sum_{k=1}^m S_{j,k}.
\end{equation}

\subsection{Consensus search engine}

From our definitions of scores in the previous subsection, we can define the \emph{consensus SE} as the one maximizing the SE score for each keyword.
Formally, for a given keyword $k$, the goal of the consensus SE is to find an ordered list  of  the $\ell$ pages (actually, getting the first $a$ is sufficient), where $\pi^{(k)}(p)$ is for the page at position $p$,
such that 
$$\pi^{(k)}(\cdot)=\mbox{argmax}_{\pi(\cdot)}
\sum_{p=1}^a q_p R_{\pi(p),k}.
$$
Note that this maximization is easy to solve: it suffices to order the pages such that $R_{\pi^{(k)}(1),k}\geq R_{\pi^{(k)}(2),k}\geq \cdots$, i.e., to display pages in the decreasing order of their score (visibility).

The total score of the consensus SE can then also be computed, and is straightforwardly maximal.

\section{Analysis in practice} \label{sec:numer}

We have implement in Python a web crawler that looks, for a set of keywords, the results provided by nine different search engines. From those results, the scores can be computed as described in the previous section, as well as the results and score of a consensus SE. The brute results can be found at \url{https://partage.mines-telecom.fr/index.php/s/aG3SYhVYPtRCBKH}. The code to get page URLs is adapted to each SE, because they display the results differently. It also deals with results display that can group pages and subpages (that is, lower level pages on a same web site) that could be treated as different otherwise. \hbruno{Probably say exactly what we do(?)}\hpat{I'm not sure what we do exactly :-)} Another solved issue is  that \emph{a priori} different URLs can lead to the same page. It is for example the case of \url{http://www.maps.com/FunFacts.aspx}, \url{http://www.maps.com/funfacts.aspx}, \url{http://www.maps.com/FunFacts.aspx?nav=FF}, 
\url{http://www.maps.com/FunFacts}, etc. It can be checked that they actually lead to the same web page output when accessing the links proposed by the SEs. Note on the other hand that it requires a longer time for our crawler to get to each page and check whether the URL gets modified.

We (arbitrarily) consider the following set of nine SEs among the most popular, in terms of number of requests according to \url{https://www.alexa.com/siteinfo}:
\begin{itemize}
\item Google
\item Yahoo!
\item Bing
\item AOL
\item ask.com
\item duckduckgo
\item Ecosia
\item StartPage
\item Qwant.
\end{itemize}
We include SEs such as Qwant or StartPage, which are said to respect privacy and neutrality.
We also clear the cookies to prevent them from affecting the results (most SEs use cookies to learn our preferences).

We consider 216 different queries included in February 2018 common searches. The choice is based on the so-called trending searches in various domains according to \url{https://trends.google.fr/trends/topcharts}. We arbitrarily chose keywords in different categories to cover a large range of possibilities.

We limit ourselves to the first page of search engines results, usually made of the first 10 links. That is, we let $a=10$. 
The click-through rates $q_p$ are set as measured in \cite{rmDEJ12a} and displayed in \cref{tab:CTRs}.
\begin{table}[bht]
\begin{center}
\begin{tabular}{cccccccccc}
$q_1$&$q_2$&$q_3$&$q_4$&$q_5$&$q_6$&$q_7$&$q_8$&$q_9$&$q_{10}$\\
\hline
0.364&0.125&0.095&0.079&0.061&0.041&0.038&0.035& 0.03&0.022
\end{tabular}
\caption{CTR values used in the simulations, taken from \cite{rmDEJ12a}}\label{tab:CTRs} 
\end{center}
\end{table}

\subsection{Search engines scores}

\cref{tab:Scores-SE} provides the average scores of the nine considered search engines, as well as that of the consensus SE, according to \cref{eqn:SEscore}. We also include the 95\% confidence intervals that would be obtained (abusively) assuming requests are independently drawn from a distribution on all possible requests.
\begin{table}[bht]
\begin{center}
{\small
\begin{tabular}{l|l} 
SE & Score \\
\hline
Google & $  0.0832\pm   0.0045$\\
Yahoo & $  0.1103\pm   0.0030$\\
Bing & $  0.0933\pm   0.0045$\\
AOL & $  0.1055\pm   0.0036$\\
Ask & $  0.0211\pm   0.0006$\\
DuckDuckGo & $  0.1106\pm   0.0029$\\
Ecosia & $  0.1071\pm   0.0033$\\
StartPage & $  0.0816\pm   0.0046$\\
Qwant & $  0.0906\pm   0.0048$\\
Consensus & $  0.1332\pm   0.0026$
\end{tabular}
}
\caption{SE scores, and 95\% confidence intervals half-widths.}\label{tab:Scores-SE} 
\end{center}
\end{table}
Under the same assumption, we can also implement statistical tests to determine whether the scores of search engines are significantly different. The corresponding $p$-values are given in Table~\ref{tab:compare_tests}.  For two search engines, the $p$-value is the probability of error when rejecting the hypothesis they have similar scores. \hpat{Schematizing, values below 1\% indicate a statistically significant difference between both search engines.}%
A  small value  indicates a statistically significant difference between both search engines (1\% means 1\% chance of error).
\begin{table}[htbp]
{\scriptsize
\begin{tabular}{cccccccccc}
&Yahoo & Bing & AOL & Ask & DuckDuckGo & Ecosia & StartPage & Qwant & Consensus\\
Google &    1.3e-38 &    5.5e-05 &    9.7e-24 &    7.7e-70 &    6.3e-42 &    5.4e-30 &    1.3e-01 &    5.4e-03 &    6.8e-82\\
Yahoo &  &    5.0e-23 &    3.5e-08 &   1.5e-131 &    5.4e-01 &    1.9e-04 &    1.0e-40 &    2.4e-25 &   2.4e-129\\
Bing &  &  &    5.8e-11 &    2.6e-81 &    3.1e-23 &    2.8e-15 &    6.6e-06 &    6.1e-02 &    4.0e-70\\
AOL &  &  &  &   6.1e-112 &    3.9e-07 &    1.3e-01 &    2.0e-26 &    1.9e-13 &    1.6e-78\\
Ask &  &  &  &  &   4.5e-135 &   5.1e-120 &    4.5e-67 &    8.3e-75 &   4.0e-163\\
DuckDuckGo &  &  &  &  &  &    4.5e-05 &    4.5e-42 &    4.4e-25 &   1.4e-130\\
Ecosia &  &  &  &  &  &  &    5.6e-32 &    1.8e-17 &    2.0e-91\\
StartPage &  &  &  &  &  &  &  &    9.4e-04 &    2.3e-82\\
Qwant &  &  &  &  &  &  &  &  &    1.6e-67\\
\end{tabular}
}
\caption{$p$-values for the tests comparing the average scores of search engines (T-test on related samples of scores).}
\label{tab:compare_tests}
\end{table}

We can remark a group of four SEs with scores above the others: DuckDuckGo, Yahoo!, Ecosia, and AOL, around 0.11. The statistical analysis using the $p$-value allows to differentiate even more, with  DuckDuckGo and Yahoo! as a first group, and Ecosia, and AOL slightly below.
Then, Bing and Qwant get scores around 0.09 (and can not be strongly differentiated from the $p$-value in Table~\ref{tab:compare_tests})), and Google and StartPage around 0.082 (since StartPage is based on Google, close results were expected). Finally, quite far from the others, Ask.com has a score around 0.02.

The consensus SE has a score of 0.133, significantly above all the SEs as shown in~\cref{tab:compare_tests}.

\subsection{Analysis}

Figure~\ref{fig:SameThanConsensus}  displays by SE the percentage  of common results with the consensus SE for each position range. Again for all our searched keywords, we count the proportion of links in the \nth{1} position correspond to the link in \nth{1} position in the consensus SE, then do the same for the first 2 positions, then for the first 3, etc.
\begin{center}
\begin{figure}[htbp]
\begin{tikzpicture} 
\begin{axis}[ xlabel=Position range $x$ (from \nth{1} to $x$), ylabel=\% in common with the consensus SE, xmin=0, xmax=11, ymin=0, ymax=103, xtick=data, legend style={at={(1.45,1)}}] 
\addplot coordinates {(1,45.0704225352)(2,46.9483568075)(3,45.8528951487)(4,44.014084507)(5,42.5352112676)(6,40.9233176839)(7,40.7109322602)(8,40.6690140845)(9,42.3056859677)(10,43.0046948357)};
\addlegendentry{Google}
\addplot coordinates {(1,90.1408450704)(2,69.7183098592)(3,67.2926447574)(4,67.0187793427)(5,66.5727699531)(6,64.9452269171)(7,63.9168343394)(8,63.9671361502)(9,63.5889410537)(10,63.1455399061)};
\addlegendentry{Yahoo}
\addplot coordinates {(1,58.6854460094)(2,56.338028169)(3,53.2081377152)(4,51.6431924883)(5,49.1079812207)(6,47.730829421)(7,46.8142186452)(8,47.6525821596)(9,47.730829421)(10,48.4976525822)};
\addlegendentry{Bing}
\addplot coordinates {(1,85.9154929577)(2,65.4929577465)(3,63.8497652582)(4,63.6150234742)(5,62.5352112676)(6,60.5633802817)(7,58.8195841717)(8,58.0399061033)(9,56.4945226917)(10,55.6807511737)};
\addlegendentry{AOL}
\addplot coordinates {(1,0.0)(2,1.40845070423)(3,4.38184663537)(4,8.21596244131)(5,12.4882629108)(6,15.3364632238)(7,15.6941649899)(8,15.0234741784)(9,14.2931664058)(10,13.8497652582)};
\addlegendentry{Ask}
\addplot coordinates {(1,92.9577464789)(2,71.3615023474)(3,68.544600939)(4,68.1924882629)(5,66.5727699531)(6,64.6322378717)(7,63.1790744467)(8,63.0868544601)(9,63.0672926448)(10,62.8169014085)};
\addlegendentry{DuckDuckGo}
\addplot coordinates {(1,87.323943662)(2,66.9014084507)(3,62.2848200313)(4,64.0845070423)(5,62.8169014085)(6,61.5023474178)(7,60.3621730382)(8,61.0328638498)(9,60.3547209181)(10,60.1877934272)};
\addlegendentry{Ecosia}
\addplot coordinates {(1,42.7230046948)(2,45.7746478873)(3,45.5399061033)(4,42.4882629108)(5,41.220657277)(6,39.4366197183)(7,40.1073105298)(8,40.6103286385)(9,42.2013562859)(10,43.3802816901)};
\addlegendentry{StartPage}
\addplot coordinates {(1,57.7464788732)(2,55.1643192488)(3,47.730829421)(4,44.4835680751)(5,42.9107981221)(6,44.6009389671)(7,45.405767941)(8,46.6549295775)(9,47.3135106938)(10,47.9812206573)};
\addlegendentry{Qwant}
\addplot coordinates {(1,100.0)(2,100.0)(3,100.0)(4,100.0)(5,100.0)(6,100.0)(7,100.0)(8,100.0)(9,100.0)(10,100.0)};
\addlegendentry{Consensus}
\end{axis} 
\end{tikzpicture}
\caption{Similarities in position with the consensus}
\label{fig:SameThanConsensus}
\end{figure}
\end{center}
The results are consistent with the previous tables: the figure highlights the same  groups of SE, while Ask.com clearly is far from the consensus.

We also draw in Figure~\ref{fig:distribrelativescore} the distribution of the score of SEs relatively to the consensus where on the $x$-axis, we have the pages ordered (for each SE) by the relative score from the largest to the smallest.
\begin{center}
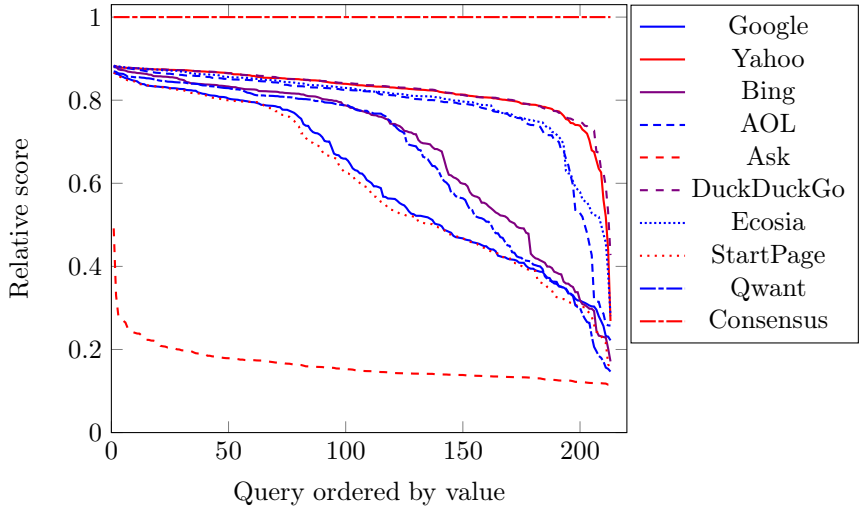
\begin{figure}[htbp]
\begin{tikzpicture} 
\begin{axis}[xlabel=Query ordered by value, ylabel=Relative score, cycle list name = \mylist,xmin=0, xmax=220, ymin=0, ymax=1.03,  legend style={at={(1.45,1)}}] 
\addplot coordinates {(1,0.866715761562)(2,0.863531370598)(3,0.862565560904)(4,0.858101969845)(5,0.855932328246)(6,0.853136311637)(7,0.852047011132)(8,0.850026224596)(9,0.84785122231)(10,0.84662113945)(11,0.845388285184)(12,0.841076680883)(13,0.839020392796)(14,0.838504034126)(15,0.835245403125)(16,0.834719557542)(17,0.834152752853)(18,0.833359112744)(19,0.832934341214)(20,0.832571697499)(21,0.831925417988)(22,0.831867681532)(23,0.831514102794)(24,0.830458696216)(25,0.828983147987)(26,0.828900732657)(27,0.828792110571)(28,0.828147974228)(29,0.82719241962)(30,0.826887414969)(31,0.826710652049)(32,0.82459901732)(33,0.824159207008)(34,0.823887548252)(35,0.82280720278)(36,0.822725401215)(37,0.818708931102)(38,0.817132483909)(39,0.815532592074)(40,0.814890856733)(41,0.813229130069)(42,0.811848072362)(43,0.811131262242)(44,0.810812247108)(45,0.80994861926)(46,0.807957140458)(47,0.807835552065)(48,0.807573520037)(49,0.806442329383)(50,0.803779196858)(51,0.802568878387)(52,0.801872426722)(53,0.801421004528)(54,0.801004919893)(55,0.799605348409)(56,0.799236198328)(57,0.798122686205)(58,0.797358414034)(59,0.795977832373)(60,0.795382822491)(61,0.795118851033)(62,0.793594558875)(63,0.791821246024)(64,0.788810385346)(65,0.788673825063)(66,0.787447645887)(67,0.786960280144)(68,0.786622592503)(69,0.786260175462)(70,0.785478851913)(71,0.783689419676)(72,0.783089798998)(73,0.777302907094)(74,0.776967023836)(75,0.776883450232)(76,0.774467051666)(77,0.772098304194)(78,0.769527286732)(79,0.764962830785)(80,0.75957395162)(81,0.755263753177)(82,0.742319936719)(83,0.738625811955)(84,0.738112300003)(85,0.72888204099)(86,0.723370732978)(87,0.720950099229)(88,0.720185008747)(89,0.716789765167)(90,0.711051103509)(91,0.701859892772)(92,0.700424168467)(93,0.700412935763)(94,0.686534892063)(95,0.671443756135)(96,0.668436487669)(97,0.667469153178)(98,0.661333031524)(99,0.65942133731)(100,0.659275269122)(101,0.654156548345)(102,0.645642245901)(103,0.635928998942)(104,0.631700417386)(105,0.629997144405)(106,0.624110078762)(107,0.623262347937)(108,0.617206741564)(109,0.607982974285)(110,0.604449482301)(111,0.600481526945)(112,0.592715857616)(113,0.592683516588)(114,0.590867698086)(115,0.588374840351)(116,0.567765535247)(117,0.567125940966)(118,0.563464385767)(119,0.563242233211)(120,0.562718008783)(121,0.562258153503)(122,0.557930287027)(123,0.552566707734)(124,0.549156889281)(125,0.54676818554)(126,0.54031733743)(127,0.532261780384)(128,0.526769817124)(129,0.525651166924)(130,0.521840176924)(131,0.521798981103)(132,0.521715350665)(133,0.520420459471)(134,0.519225129416)(135,0.518072686444)(136,0.514836555674)(137,0.511671977521)(138,0.503006523035)(139,0.501186056197)(140,0.49996092218)(141,0.496482180803)(142,0.487858263174)(143,0.485898451525)(144,0.4833291068)(145,0.480888823587)(146,0.473348804234)(147,0.470289868525)(148,0.469352957071)(149,0.468363901481)(150,0.467465356261)(151,0.464674267327)(152,0.463376676474)(153,0.462478258488)(154,0.458318897963)(155,0.45711528169)(156,0.455807157043)(157,0.454212246335)(158,0.451343733466)(159,0.448804957272)(160,0.448357426075)(161,0.443216224616)(162,0.440827286375)(163,0.440653212151)(164,0.436394820772)(165,0.425774944082)(166,0.424116950524)(167,0.422191216317)(168,0.422004249882)(169,0.418632894861)(170,0.416982184377)(171,0.416448748882)(172,0.415737055017)(173,0.413692504064)(174,0.408883111211)(175,0.40864810971)(176,0.40856435619)(177,0.401180679444)(178,0.39647438966)(179,0.394493328381)(180,0.394447136697)(181,0.390479159778)(182,0.388782454484)(183,0.387080047435)(184,0.387040417257)(185,0.379464612982)(186,0.372973875096)(187,0.371926608703)(188,0.363297497535)(189,0.363248652259)(190,0.362711206663)(191,0.351919169218)(192,0.350392158857)(193,0.349173982082)(194,0.344909403738)(195,0.334901179735)(196,0.334443735411)(197,0.33032625252)(198,0.321939028591)(199,0.31800769434)(200,0.317732176288)(201,0.314727821311)(202,0.312348760552)(203,0.309319215534)(204,0.305863245201)(205,0.303839782836)(206,0.30333383251)(207,0.295732543698)(208,0.279996584287)(209,0.2685966679)(210,0.256372846119)(211,0.232192015922)(212,0.230893655023)(213,0.221930579623)};
\addlegendentry{Google}
\addplot coordinates {(1,0.879944861672)(2,0.879258946425)(3,0.879202633396)(4,0.878230774508)(5,0.87815130588)(6,0.877997404513)(7,0.877905258318)(8,0.877343693336)(9,0.877121627434)(10,0.876835352731)(11,0.876717175751)(12,0.876159608857)(13,0.875701631472)(14,0.875110108422)(15,0.874902522843)(16,0.874821747129)(17,0.87470076285)(18,0.874393088307)(19,0.874201191384)(20,0.874101938748)(21,0.873834150042)(22,0.873789713906)(23,0.873629556153)(24,0.873218366256)(25,0.87289423155)(26,0.872872075686)(27,0.872790888535)(28,0.8723715252)(29,0.872340141324)(30,0.872253970044)(31,0.872181468784)(32,0.872083485104)(33,0.871594534789)(34,0.871049145563)(35,0.870452753709)(36,0.869884860791)(37,0.86940893152)(38,0.869276332817)(39,0.869161175865)(40,0.8689073502)(41,0.868430093479)(42,0.868127933781)(43,0.867907104379)(44,0.867870349801)(45,0.867325058378)(46,0.866752661109)(47,0.866405220731)(48,0.866133099509)(49,0.865800690742)(50,0.86524891103)(51,0.86515939243)(52,0.862450101505)(53,0.862334632904)(54,0.862268168077)(55,0.861855690546)(56,0.860721336662)(57,0.860642769995)(58,0.860382433779)(59,0.859989337704)(60,0.858407074711)(61,0.858260356145)(62,0.856362187079)(63,0.855748365888)(64,0.855449063091)(65,0.854086978897)(66,0.853986205964)(67,0.853852547988)(68,0.853609422164)(69,0.853461086869)(70,0.853303618852)(71,0.853243592575)(72,0.853228985065)(73,0.852690554763)(74,0.852460082389)(75,0.851365597269)(76,0.851163978452)(77,0.851026046701)(78,0.850930898756)(79,0.850732763177)(80,0.850525615925)(81,0.849448370982)(82,0.8492893392)(83,0.84852149335)(84,0.848484243153)(85,0.848035490339)(86,0.847845544458)(87,0.847720952333)(88,0.847204551905)(89,0.847040880963)(90,0.845539598993)(91,0.84470666447)(92,0.843953719208)(93,0.842854060128)(94,0.842094293662)(95,0.841781006621)(96,0.840765821466)(97,0.840673954445)(98,0.840255642445)(99,0.8394994857)(100,0.839394447104)(101,0.83885519699)(102,0.838140526904)(103,0.837504775192)(104,0.837348531919)(105,0.835972165832)(106,0.835945562778)(107,0.835813036508)(108,0.835133490408)(109,0.834065112824)(110,0.834030748429)(111,0.83394404603)(112,0.833365263159)(113,0.832694743028)(114,0.832689344467)(115,0.832523832223)(116,0.832518274496)(117,0.832224144577)(118,0.832193628394)(119,0.830737990941)(120,0.830653072051)(121,0.8302247508)(122,0.829320745589)(123,0.828932746335)(124,0.82835613326)(125,0.828325124223)(126,0.828044471975)(127,0.827132541283)(128,0.826748061054)(129,0.826336556575)(130,0.825698845758)(131,0.825394905191)(132,0.825069137887)(133,0.824937601948)(134,0.824516419135)(135,0.823799400793)(136,0.823656459069)(137,0.823470251421)(138,0.822735995516)(139,0.822285628044)(140,0.8202231177)(141,0.819317077465)(142,0.818382525278)(143,0.817531314453)(144,0.817107796307)(145,0.817095452446)(146,0.816233232871)(147,0.813875988104)(148,0.813797088658)(149,0.813611821453)(150,0.81344234725)(151,0.813136559735)(152,0.81131180585)(153,0.810354984784)(154,0.809943477769)(155,0.809157100518)(156,0.808723105874)(157,0.80747546973)(158,0.807467486275)(159,0.806870369643)(160,0.80673320912)(161,0.806338592146)(162,0.806100104914)(163,0.804488086323)(164,0.804410618242)(165,0.803205230144)(166,0.802247394936)(167,0.801275651435)(168,0.801030625134)(169,0.800087045836)(170,0.799168495284)(171,0.798608748826)(172,0.797656571562)(173,0.796771158163)(174,0.795130819563)(175,0.791886877961)(176,0.791134002783)(177,0.791080504045)(178,0.790888775737)(179,0.789283495828)(180,0.789131590714)(181,0.78880472085)(182,0.787153333296)(183,0.786176409123)(184,0.784086928037)(185,0.783729816306)(186,0.781701096278)(187,0.780520224378)(188,0.778903315521)(189,0.774104033867)(190,0.774071365709)(191,0.773003067747)(192,0.768243820629)(193,0.762432147176)(194,0.754582797509)(195,0.752747696692)(196,0.745136321283)(197,0.74451292864)(198,0.740429069903)(199,0.740209221937)(200,0.738764703132)(201,0.729625687661)(202,0.7250109717)(203,0.719626553616)(204,0.703744541518)(205,0.679357930135)(206,0.675599614157)(207,0.655188224401)(208,0.63506392953)(209,0.626413856717)(210,0.560730145058)(211,0.519453344678)(212,0.44004865493)(213,0.268551809254)};
\addlegendentry{Yahoo}
\addplot coordinates {(1,0.883463471962)(2,0.881464972541)(3,0.874175111965)(4,0.87276041753)(5,0.872380855607)(6,0.871292612566)(7,0.868804377511)(8,0.868058245296)(9,0.866976985156)(10,0.866788852097)(11,0.865425856917)(12,0.865201106002)(13,0.864437664141)(14,0.863836589644)(15,0.863146895847)(16,0.861477882387)(17,0.861173541696)(18,0.857842339225)(19,0.857826003054)(20,0.857510989367)(21,0.857476655918)(22,0.856678454416)(23,0.856582602578)(24,0.856186136695)(25,0.855276119915)(26,0.855242024202)(27,0.854166690557)(28,0.853436335082)(29,0.850298465366)(30,0.849404896871)(31,0.848122058941)(32,0.84378622587)(33,0.842801376842)(34,0.840685860884)(35,0.840526407975)(36,0.839918749298)(37,0.839159766873)(38,0.839019294885)(39,0.838502936115)(40,0.837834111128)(41,0.837742677828)(42,0.837426854237)(43,0.836964956962)(44,0.836769631169)(45,0.836354341567)(46,0.834847232867)(47,0.834748462924)(48,0.834212996672)(49,0.833442038839)(50,0.832898017989)(51,0.831752979953)(52,0.831364341875)(53,0.83124402485)(54,0.829310188911)(55,0.828606261267)(56,0.827570416897)(57,0.826570327171)(58,0.826330620226)(59,0.826228210177)(60,0.825968159014)(61,0.825801595277)(62,0.825029068221)(63,0.821930932168)(64,0.821821199424)(65,0.821709895948)(66,0.821627806843)(67,0.821180710305)(68,0.820855710658)(69,0.82061168778)(70,0.820512334035)(71,0.820032346982)(72,0.818482468991)(73,0.818438921564)(74,0.817482797255)(75,0.817168368843)(76,0.816851620474)(77,0.816207181354)(78,0.816115593999)(79,0.815656227582)(80,0.815561672251)(81,0.813906668452)(82,0.813668984852)(83,0.811146506691)(84,0.809844838221)(85,0.809457851712)(86,0.809261294301)(87,0.808854812756)(88,0.80742044284)(89,0.805420345235)(90,0.802921866154)(91,0.802004877666)(92,0.801455640408)(93,0.800302555575)(94,0.800039828379)(95,0.79848156957)(96,0.798419147119)(97,0.795771738034)(98,0.795126319581)(99,0.788065735822)(100,0.787356920639)(101,0.785616234559)(102,0.7816723656)(103,0.781514901434)(104,0.778889478148)(105,0.77782724954)(106,0.774640428661)(107,0.773750624115)(108,0.772299784112)(109,0.769532952414)(110,0.768791552118)(111,0.76467579133)(112,0.763903092286)(113,0.763536436585)(114,0.760673645361)(115,0.760123858451)(116,0.758225387472)(117,0.755112566033)(118,0.753577074044)(119,0.743531691354)(120,0.742872226347)(121,0.742739713024)(122,0.739680065)(123,0.737724857916)(124,0.733314503309)(125,0.726118559352)(126,0.725355967214)(127,0.722974927093)(128,0.722247864211)(129,0.71851944306)(130,0.718011436191)(131,0.704481837186)(132,0.698652085301)(133,0.696385478342)(134,0.695826836443)(135,0.694969791504)(136,0.693734365667)(137,0.687147782095)(138,0.682534949057)(139,0.682321553332)(140,0.67956949297)(141,0.676513092507)(142,0.657085262723)(143,0.636506967471)(144,0.620529623718)(145,0.61897794825)(146,0.614356673239)(147,0.612090470329)(148,0.608174306929)(149,0.600953592677)(150,0.599109158626)(151,0.598101884603)(152,0.592546543953)(153,0.57880633237)(154,0.571686042455)(155,0.568944151053)(156,0.567535737143)(157,0.564493327226)(158,0.553330946808)(159,0.553233954901)(160,0.550338537603)(161,0.544463253332)(162,0.540102419892)(163,0.535243936636)(164,0.529982635753)(165,0.526999028755)(166,0.522730583815)(167,0.521797385799)(168,0.520439334936)(169,0.516396752242)(170,0.508901892617)(171,0.506116036247)(172,0.503449502779)(173,0.495719048545)(174,0.4928457915)(175,0.491795667831)(176,0.491347618649)(177,0.480973167961)(178,0.478770195324)(179,0.430467874195)(180,0.424133382442)(181,0.419458395948)(182,0.417845011446)(183,0.41452565966)(184,0.411759873403)(185,0.408690577795)(186,0.399530912923)(187,0.399498332009)(188,0.396071279723)(189,0.387112336017)(190,0.384901282854)(191,0.382026258013)(192,0.380794897148)(193,0.363367360116)(194,0.358848838237)(195,0.358337639946)(196,0.353331044601)(197,0.350262052988)(198,0.337885899817)(199,0.337843250272)(200,0.318057502025)(201,0.309337974326)(202,0.307234305668)(203,0.30298826678)(204,0.298138157221)(205,0.298108171391)(206,0.292792583217)(207,0.245215353346)(208,0.233239943755)(209,0.231808958638)(210,0.230107859037)(211,0.228713558893)(212,0.202469269583)(213,0.17134421104)};
\addlegendentry{Bing}
\addplot coordinates {(1,0.879944861672)(2,0.879258946425)(3,0.879202633396)(4,0.878950643884)(5,0.878844633378)(6,0.878258994641)(7,0.876562818772)(8,0.876159608857)(9,0.875110108422)(10,0.874495852655)(11,0.874201191384)(12,0.872747707418)(13,0.871857991612)(14,0.870810608295)(15,0.869623362352)(16,0.868220099235)(17,0.867475875096)(18,0.867139855891)(19,0.866133099509)(20,0.865741738637)(21,0.865441565556)(22,0.86524891103)(23,0.864645566258)(24,0.863743728678)(25,0.863681938378)(26,0.862921553403)(27,0.862115164901)(28,0.861606380915)(29,0.861140748846)(30,0.860877805059)(31,0.859989337704)(32,0.859355465417)(33,0.859289115016)(34,0.858964974374)(35,0.858929275216)(36,0.858394453784)(37,0.858260356145)(38,0.858244446723)(39,0.857111217138)(40,0.856840006275)(41,0.855921835373)(42,0.855748365888)(43,0.853946717917)(44,0.853478158731)(45,0.853303618852)(46,0.853050998898)(47,0.852656217862)(48,0.851455647107)(49,0.851365597269)(50,0.850778965466)(51,0.850743215376)(52,0.84980966882)(53,0.849771599329)(54,0.849600396538)(55,0.849352137393)(56,0.849117177734)(57,0.849056439569)(58,0.848918403846)(59,0.848381118817)(60,0.848209652517)(61,0.847943545478)(62,0.846130903101)(63,0.8460611914)(64,0.845585937905)(65,0.845304760127)(66,0.84526663423)(67,0.844950837968)(68,0.844115645205)(69,0.8429333612)(70,0.842854060128)(71,0.842491478918)(72,0.840144177452)(73,0.838155118512)(74,0.837243494724)(75,0.836988466943)(76,0.835696245697)(77,0.834554440031)(78,0.834464776534)(79,0.83377718005)(80,0.833679832595)(81,0.833582423516)(82,0.833422021852)(83,0.831261506616)(84,0.831118573269)(85,0.830994423784)(86,0.830653072051)(87,0.830632603999)(88,0.830381304067)(89,0.83010827583)(90,0.829880088633)(91,0.82942316898)(92,0.829069498992)(93,0.828830328235)(94,0.827850714888)(95,0.827738035881)(96,0.827124871013)(97,0.82688010908)(98,0.825572551353)(99,0.824992896683)(100,0.824806452766)(101,0.824536987025)(102,0.82428310783)(103,0.823871797403)(104,0.823622754463)(105,0.823564519448)(106,0.823470251421)(107,0.822327969038)(108,0.821627571233)(109,0.821299709165)(110,0.821264791701)(111,0.821014011368)(112,0.82097771436)(113,0.820152860727)(114,0.816057431518)(115,0.813875988104)(116,0.813790742236)(117,0.812883568588)(118,0.812586036749)(119,0.812265662355)(120,0.811886048243)(121,0.811627303772)(122,0.808959629427)(123,0.808736720901)(124,0.808020562636)(125,0.807712255844)(126,0.806602101034)(127,0.806476156257)(128,0.806338592146)(129,0.803305945755)(130,0.802785756343)(131,0.802528343967)(132,0.801990262424)(133,0.801796909609)(134,0.801689321703)(135,0.800326198401)(136,0.800288523782)(137,0.800277611331)(138,0.79978711799)(139,0.797355704801)(140,0.79729608683)(141,0.796771158163)(142,0.795392148657)(143,0.795373220138)(144,0.795175087001)(145,0.794527060674)(146,0.794228659811)(147,0.793947208931)(148,0.79208546319)(149,0.791198913662)(150,0.791164702351)(151,0.791134002783)(152,0.790393322989)(153,0.789376257455)(154,0.788783464022)(155,0.786217267919)(156,0.786159807285)(157,0.784071132505)(158,0.781701096278)(159,0.780478920808)(160,0.779168423752)(161,0.778903315521)(162,0.778567521026)(163,0.77348953527)(164,0.772289797193)(165,0.771524766283)(166,0.770764976419)(167,0.769725059839)(168,0.768049501915)(169,0.76714453839)(170,0.76237762759)(171,0.761279488735)(172,0.758184141633)(173,0.757397174296)(174,0.756823175782)(175,0.756731030899)(176,0.745888141789)(177,0.745175775381)(178,0.742012683643)(179,0.7403530419)(180,0.740197401836)(181,0.738892760195)(182,0.738763605112)(183,0.736004972882)(184,0.726379377491)(185,0.716180492958)(186,0.716160923148)(187,0.715769522911)(188,0.715555999186)(189,0.703744541518)(190,0.703602051789)(191,0.699170706122)(192,0.679357930135)(193,0.678031596487)(194,0.659213074533)(195,0.643718040724)(196,0.618748402359)(197,0.594975340602)(198,0.536078556767)(199,0.536046332027)(200,0.527958847967)(201,0.519453344678)(202,0.499703365917)(203,0.480200210603)(204,0.453154742091)(205,0.426998430374)(206,0.319636557494)(207,0.315629433149)(208,0.314749229741)(209,0.293585249484)(210,0.28619600479)(211,0.260914058201)(212,0.258051287496)(213,0.257573010148)};
\addlegendentry{AOL}
\addplot coordinates {(1,0.491491342227)(2,0.330148847052)(3,0.280169679898)(4,0.275592105037)(5,0.269425883699)(6,0.26494922395)(7,0.250909811628)(8,0.244157529087)(9,0.24213070921)(10,0.240201622188)(11,0.238444177304)(12,0.238143830877)(13,0.232744304686)(14,0.230543468619)(15,0.223214408799)(16,0.223089893685)(17,0.22141538929)(18,0.221220679751)(19,0.219609279952)(20,0.21689205405)(21,0.208717204212)(22,0.20769021004)(23,0.206943677128)(24,0.204752028495)(25,0.204447667248)(26,0.204246785229)(27,0.202451946272)(28,0.20040295883)(29,0.199992878331)(30,0.197969748443)(31,0.19787496212)(32,0.19750344218)(33,0.19687585859)(34,0.196053594262)(35,0.193298643398)(36,0.191010298679)(37,0.190366330521)(38,0.190243788658)(39,0.189304485766)(40,0.188911351428)(41,0.184961109074)(42,0.184887277921)(43,0.183486290785)(44,0.183061187094)(45,0.18143466415)(46,0.181013635544)(47,0.180550915493)(48,0.17966671066)(49,0.179122177954)(50,0.178970158719)(51,0.17829956952)(52,0.177798807268)(53,0.177725978803)(54,0.176775860283)(55,0.175627170659)(56,0.175603312766)(57,0.174621190775)(58,0.174372350682)(59,0.174261348897)(60,0.17421047468)(61,0.174194650822)(62,0.173677422562)(63,0.173482912152)(64,0.173411530228)(65,0.172883638281)(66,0.172365474974)(67,0.171203067266)(68,0.170276974201)(69,0.169539223531)(70,0.168595357416)(71,0.16858088145)(72,0.168405808621)(73,0.167532931885)(74,0.166373739135)(75,0.165934465169)(76,0.165723458014)(77,0.164650366753)(78,0.164016790917)(79,0.163206558985)(80,0.163031877349)(81,0.162591383605)(82,0.16113432946)(83,0.159455530337)(84,0.159366241322)(85,0.159015840659)(86,0.159000390707)(87,0.15821835723)(88,0.158093639745)(89,0.157934411216)(90,0.157666373508)(91,0.156929103106)(92,0.156841640549)(93,0.156825554349)(94,0.156115219037)(95,0.15569283913)(96,0.154331462975)(97,0.154106617805)(98,0.153974060152)(99,0.153628412404)(100,0.153128978074)(101,0.152337885879)(102,0.152104343053)(103,0.151558428507)(104,0.150207868295)(105,0.149427543426)(106,0.149293398668)(107,0.148622107229)(108,0.148214379001)(109,0.147988709598)(110,0.147893098903)(111,0.147775280857)(112,0.146995556575)(113,0.14685129993)(114,0.146207162022)(115,0.14572489023)(116,0.145669486623)(117,0.14538786952)(118,0.145072649786)(119,0.144065081005)(120,0.14389366624)(121,0.143317660645)(122,0.143022264123)(123,0.142879353139)(124,0.142812872942)(125,0.142528046164)(126,0.142260213112)(127,0.142213521756)(128,0.142168061172)(129,0.141903265697)(130,0.141855198767)(131,0.1417845285)(132,0.14174478837)(133,0.141710404122)(134,0.141621858998)(135,0.141435656309)(136,0.141345637596)(137,0.141338701005)(138,0.140597470826)(139,0.140507447391)(140,0.140166023552)(141,0.139978078999)(142,0.139801552428)(143,0.139743700815)(144,0.139347041932)(145,0.139275934695)(146,0.139196740269)(147,0.138957374643)(148,0.138717120718)(149,0.13862444338)(150,0.138602567837)(151,0.138027941282)(152,0.137322158028)(153,0.137247910893)(154,0.136902452074)(155,0.136845608389)(156,0.136652562812)(157,0.136567397521)(158,0.136280656184)(159,0.135926981036)(160,0.135881105741)(161,0.135575221546)(162,0.135415231518)(163,0.13504656373)(164,0.135014962216)(165,0.134925625271)(166,0.134756013806)(167,0.134543427072)(168,0.134347278153)(169,0.133882611379)(170,0.133758729384)(171,0.133567694947)(172,0.133335451294)(173,0.1332716889)(174,0.133030235681)(175,0.133029404859)(176,0.133026573744)(177,0.13282783564)(178,0.132451450919)(179,0.132207958293)(180,0.131835749289)(181,0.130650204601)(182,0.130581403167)(183,0.130347287258)(184,0.129089979708)(185,0.128013741726)(186,0.127755082772)(187,0.127381096291)(188,0.126964251926)(189,0.12678152365)(190,0.126207173237)(191,0.125405271885)(192,0.125202383785)(193,0.125053556273)(194,0.124986191582)(195,0.124968700234)(196,0.12490622049)(197,0.123734210886)(198,0.122109233502)(199,0.122054455406)(200,0.121839649093)(201,0.121332047154)(202,0.121044863028)(203,0.120187278208)(204,0.119580367068)(205,0.119312644193)(206,0.119299934484)(207,0.119209323519)(208,0.119045303688)(209,0.118800376258)(210,0.118280068771)(211,0.117944100082)(212,0.115278774471)(213,0.115048177909)};
\addlegendentry{Ask}
\addplot coordinates {(1,0.882088877145)(2,0.881345810525)(3,0.879944861672)(4,0.879627333276)(5,0.879202633396)(6,0.878213055624)(7,0.878193006276)(8,0.87815130588)(9,0.877420520051)(10,0.877159223434)(11,0.876835352731)(12,0.876803543088)(13,0.876159608857)(14,0.876001723553)(15,0.875701631472)(16,0.875110108422)(17,0.87503740597)(18,0.874631874123)(19,0.874380427063)(20,0.874285165068)(21,0.874201191384)(22,0.874188547748)(23,0.873834150042)(24,0.873307119403)(25,0.873218366256)(26,0.872991929075)(27,0.872872075686)(28,0.872596355785)(29,0.872370399995)(30,0.872340141324)(31,0.872083485104)(32,0.871856801122)(33,0.871598867922)(34,0.871057194059)(35,0.870420331057)(36,0.869986005064)(37,0.869977158665)(38,0.869538359841)(39,0.869276332817)(40,0.869161175865)(41,0.869144735286)(42,0.868606292154)(43,0.8684018864)(44,0.868127933781)(45,0.867521141102)(46,0.867325058378)(47,0.866785324968)(48,0.865812529578)(49,0.865279543677)(50,0.864739667872)(51,0.864584277174)(52,0.862268168077)(53,0.862218743937)(54,0.86219217836)(55,0.861855690546)(56,0.861017193013)(57,0.860642769995)(58,0.860149697799)(59,0.860045113793)(60,0.859989337704)(61,0.859948456392)(62,0.859886555527)(63,0.85867847963)(64,0.858407074711)(65,0.858298954624)(66,0.857938943036)(67,0.856472909167)(68,0.856069338905)(69,0.856014127597)(70,0.855748365888)(71,0.853986205964)(72,0.853243592575)(73,0.852842969326)(74,0.8512765257)(75,0.851163978452)(76,0.851026046701)(77,0.850930898756)(78,0.85073997082)(79,0.850118729606)(80,0.850029247879)(81,0.849771599329)(82,0.849712638595)(83,0.849151139347)(84,0.848979932306)(85,0.847204551905)(86,0.847040880963)(87,0.846665832896)(88,0.846658087838)(89,0.846496818202)(90,0.846217747532)(91,0.846080343135)(92,0.845401730159)(93,0.845279745192)(94,0.843893923997)(95,0.843722700979)(96,0.843056306771)(97,0.842094293662)(98,0.841789188751)(99,0.841414919265)(100,0.841017354935)(101,0.840765821466)(102,0.840595958889)(103,0.840241476445)(104,0.8394994857)(105,0.839394447104)(106,0.839148253097)(107,0.837348531919)(108,0.837103296888)(109,0.836353878307)(110,0.836229209658)(111,0.835972165832)(112,0.835942933043)(113,0.835686697347)(114,0.835398757248)(115,0.83394404603)(116,0.833898684355)(117,0.833692367144)(118,0.832689344467)(119,0.832454557106)(120,0.832193628394)(121,0.831973717359)(122,0.831969527668)(123,0.830508670807)(124,0.83038760962)(125,0.8302247508)(126,0.829320745589)(127,0.828862511193)(128,0.828510061127)(129,0.828325124223)(130,0.828044471975)(131,0.827414221545)(132,0.827132541283)(133,0.826946275302)(134,0.826748061054)(135,0.826336556575)(136,0.825394905191)(137,0.825388425495)(138,0.824536987025)(139,0.821599672403)(140,0.820989765027)(141,0.820239195683)(142,0.819373485915)(143,0.818106680366)(144,0.817972210333)(145,0.817095452446)(146,0.816933793152)(147,0.813691843109)(148,0.813476336774)(149,0.813110252784)(150,0.812583971367)(151,0.81131180585)(152,0.810415840823)(153,0.810402368154)(154,0.809619196869)(155,0.808428456934)(156,0.80747546973)(157,0.807294998226)(158,0.80673320912)(159,0.806602101034)(160,0.806476914322)(161,0.806408744306)(162,0.806298129609)(163,0.80608134877)(164,0.805827554992)(165,0.804410618242)(166,0.804021362066)(167,0.803531429761)(168,0.803205230144)(169,0.802111612558)(170,0.800944432484)(171,0.800563321336)(172,0.799168495284)(173,0.797523721346)(174,0.795130819563)(175,0.792868530731)(176,0.79208546319)(177,0.791164702351)(178,0.7900773845)(179,0.788618328553)(180,0.787325986817)(181,0.786176409123)(182,0.78568407284)(183,0.785628632747)(184,0.781701096278)(185,0.779840276355)(186,0.776635156081)(187,0.776293046341)(188,0.775194919066)(189,0.774071365709)(190,0.773435335607)(191,0.769634004301)(192,0.769576305477)(193,0.764858979867)(194,0.758707608113)(195,0.756480347929)(196,0.75566726514)(197,0.752747696692)(198,0.750548378759)(199,0.749685263484)(200,0.747973330073)(201,0.745136321283)(202,0.738556654476)(203,0.737549098457)(204,0.729625687661)(205,0.729264557502)(206,0.726784807813)(207,0.675740834813)(208,0.673948201962)(209,0.6319191656)(210,0.626082904632)(211,0.592884205862)(212,0.530612992376)(213,0.427228285298)};
\addlegendentry{DuckDuckGo}
\addplot coordinates {(1,0.882088877145)(2,0.88176378223)(3,0.881345810525)(4,0.880647847128)(5,0.879851933239)(6,0.879009637323)(7,0.878772077376)(8,0.877438055912)(9,0.877364066864)(10,0.877002899354)(11,0.87694309073)(12,0.876537148718)(13,0.876341916442)(14,0.876130365009)(15,0.874900871236)(16,0.874285165068)(17,0.874213706591)(18,0.874011022085)(19,0.873798171246)(20,0.873476628954)(21,0.873143704269)(22,0.872053947043)(23,0.871934917219)(24,0.871276213291)(25,0.871231397618)(26,0.87119657321)(27,0.871049972806)(28,0.870828734758)(29,0.869113967142)(30,0.869060496354)(31,0.868541022366)(32,0.867651049273)(33,0.865195340913)(34,0.865146862495)(35,0.864834520403)(36,0.86330274244)(37,0.863108525824)(38,0.862684407542)(39,0.862341584934)(40,0.862277015194)(41,0.862173187397)(42,0.861663583998)(43,0.861507213229)(44,0.861037608906)(45,0.86074972998)(46,0.859630481176)(47,0.858144414962)(48,0.856673933805)(49,0.856553568427)(50,0.855823630833)(51,0.85574946762)(52,0.854888723198)(53,0.854828789456)(54,0.85479031776)(55,0.854191252051)(56,0.85409197499)(57,0.853935372812)(58,0.853556946047)(59,0.852555085013)(60,0.852246422379)(61,0.852037279341)(62,0.851275294336)(63,0.850868389306)(64,0.850672935888)(65,0.850458477571)(66,0.848504831124)(67,0.848405139568)(68,0.848331721873)(69,0.847927319105)(70,0.847766697983)(71,0.846706168937)(72,0.845901119723)(73,0.844623441088)(74,0.844027297423)(75,0.843601150592)(76,0.843239483336)(77,0.84298719274)(78,0.842367036094)(79,0.841483277319)(80,0.840279089241)(81,0.840142952543)(82,0.840006644772)(83,0.83826490802)(84,0.836863229508)(85,0.836807906201)(86,0.836072547277)(87,0.835980387291)(88,0.835913286901)(89,0.835881556399)(90,0.83472221505)(91,0.834096878021)(92,0.83376585244)(93,0.833291555192)(94,0.833283922938)(95,0.832351867515)(96,0.832113967033)(97,0.830778030106)(98,0.830545620595)(99,0.829650817338)(100,0.829462142921)(101,0.829212052878)(102,0.829142096561)(103,0.827924844423)(104,0.827267418017)(105,0.825685659634)(106,0.824465769648)(107,0.823983434545)(108,0.823926761407)(109,0.822769255617)(110,0.821860807568)(111,0.820792371785)(112,0.820681864933)(113,0.820657405639)(114,0.820248303735)(115,0.819267730193)(116,0.819265221594)(117,0.819115235733)(118,0.817926593707)(119,0.816986348772)(120,0.815727259653)(121,0.815483490566)(122,0.815315402735)(123,0.81524825757)(124,0.814713459521)(125,0.813579241662)(126,0.8128548252)(127,0.811489400553)(128,0.810691547118)(129,0.810255470275)(130,0.810210729326)(131,0.81003682933)(132,0.809720124928)(133,0.809408275636)(134,0.809381470681)(135,0.809242700333)(136,0.809149171409)(137,0.808571190812)(138,0.807352469311)(139,0.807248775415)(140,0.807008514176)(141,0.803640120347)(142,0.803222802708)(143,0.802880628759)(144,0.801827899991)(145,0.80109400875)(146,0.799266974323)(147,0.799200155769)(148,0.798936262311)(149,0.797701420413)(150,0.796828910794)(151,0.795938474618)(152,0.795913960428)(153,0.795502791375)(154,0.795113438493)(155,0.79298978147)(156,0.791708711832)(157,0.788812158201)(158,0.788782839975)(159,0.788307069692)(160,0.788151733596)(161,0.788038167249)(162,0.787481069409)(163,0.777188238934)(164,0.776111662642)(165,0.774190509363)(166,0.769503017456)(167,0.769244834553)(168,0.76877049068)(169,0.768624140141)(170,0.767624781088)(171,0.765132802349)(172,0.75981381207)(173,0.758823548358)(174,0.758641065405)(175,0.757037424586)(176,0.756684277334)(177,0.753714852168)(178,0.753119962108)(179,0.7514313412)(180,0.751407907898)(181,0.748518452931)(182,0.748478964321)(183,0.747358096063)(184,0.746533639792)(185,0.739400611071)(186,0.734107304141)(187,0.733695881317)(188,0.723965260454)(189,0.722307077465)(190,0.717076608969)(191,0.711356172422)(192,0.705028955388)(193,0.693502581316)(194,0.629458055321)(195,0.623817816986)(196,0.623430912211)(197,0.604695117179)(198,0.590982575529)(199,0.584736443769)(200,0.579402108211)(201,0.56203030427)(202,0.557020045814)(203,0.554494393688)(204,0.543797058429)(205,0.528005889556)(206,0.523834517101)(207,0.521916480726)(208,0.51935120255)(209,0.500546714504)(210,0.477878735719)(211,0.457152121253)(212,0.382208267268)(213,0.28027583787)};
\addlegendentry{Ecosia}
\addplot coordinates {(1,0.866715761562)(2,0.863531370598)(3,0.856764171362)(4,0.85570422778)(5,0.851303379799)(6,0.850026224596)(7,0.8478523919)(8,0.846635542615)(9,0.846082566737)(10,0.844944100248)(11,0.839272237218)(12,0.838504034126)(13,0.838323335144)(14,0.838091164066)(15,0.836287512638)(16,0.835095086796)(17,0.834609020749)(18,0.833967343144)(19,0.833359112744)(20,0.831879634798)(21,0.8312891775)(22,0.830982700512)(23,0.830474665324)(24,0.828548840903)(25,0.828437633641)(26,0.827264659776)(27,0.825694037795)(28,0.825383375566)(29,0.82533854549)(30,0.824961085754)(31,0.823462314232)(32,0.821589956254)(33,0.819151608756)(34,0.817602151493)(35,0.816834509374)(36,0.816690494176)(37,0.816232480502)(38,0.815852447126)(39,0.814058449489)(40,0.812278621896)(41,0.809938598087)(42,0.807302695617)(43,0.806161785142)(44,0.805068520633)(45,0.803830528134)(46,0.803668402914)(47,0.802568878387)(48,0.801522232194)(49,0.80015064107)(50,0.79979140006)(51,0.798936207)(52,0.798122686205)(53,0.798022071395)(54,0.797142466151)(55,0.796813860191)(56,0.796791331725)(57,0.796741289052)(58,0.796348189933)(59,0.795977832373)(60,0.795636305972)(61,0.795457467555)(62,0.795382822491)(63,0.794216016696)(64,0.793071997827)(65,0.787447645887)(66,0.783089798998)(67,0.781095501134)(68,0.780804430126)(69,0.779383386471)(70,0.77355653583)(71,0.773486202804)(72,0.766855904417)(73,0.764627839136)(74,0.764406070226)(75,0.760250473676)(76,0.76011718423)(77,0.754978703501)(78,0.750315419508)(79,0.742184195943)(80,0.736893783972)(81,0.720281396232)(82,0.720185008747)(83,0.709569271608)(84,0.701651399715)(85,0.699167490642)(86,0.697323211626)(87,0.691592300474)(88,0.6907203571)(89,0.688208127876)(90,0.676651551364)(91,0.668436487669)(92,0.667333614109)(93,0.666017431367)(94,0.664113193545)(95,0.661333031524)(96,0.65942133731)(97,0.645207434225)(98,0.635928998942)(99,0.629997144405)(100,0.624110078762)(101,0.623754285698)(102,0.621019865574)(103,0.61944760284)(104,0.607982974285)(105,0.60567014934)(106,0.60077475594)(107,0.594063022862)(108,0.592715857616)(109,0.590867698086)(110,0.584691433453)(111,0.574886410822)(112,0.567765535247)(113,0.567477692376)(114,0.563467856242)(115,0.555495572286)(116,0.552649415857)(117,0.548787993388)(118,0.54031733743)(119,0.536731345407)(120,0.53560484834)(121,0.535095750852)(122,0.531224059889)(123,0.526890985759)(124,0.525744468882)(125,0.524400120291)(126,0.52204504305)(127,0.521798981103)(128,0.520667498471)(129,0.520046939102)(130,0.520018149548)(131,0.507456948297)(132,0.505747714675)(133,0.504709649495)(134,0.50348512406)(135,0.502545334718)(136,0.502273967018)(137,0.501372106015)(138,0.496482180803)(139,0.495297917923)(140,0.492447004899)(141,0.489449623266)(142,0.487506978536)(143,0.48021347597)(144,0.477106134961)(145,0.476895513934)(146,0.476446461832)(147,0.473348804234)(148,0.470677335681)(149,0.468871608902)(150,0.468350227956)(151,0.463901043378)(152,0.462138343794)(153,0.45941681102)(154,0.457387314863)(155,0.454815583044)(156,0.454507858048)(157,0.448357426075)(158,0.446539220741)(159,0.444699969449)(160,0.442434099)(161,0.437789860244)(162,0.436394820772)(163,0.433388820329)(164,0.433362032416)(165,0.430594853383)(166,0.424116950524)(167,0.4227782272)(168,0.422004249882)(169,0.421998063665)(170,0.419559835238)(171,0.408604081607)(172,0.40856435619)(173,0.405726823041)(174,0.401180679444)(175,0.401119059832)(176,0.397724489213)(177,0.394106347323)(178,0.392031059548)(179,0.390479159778)(180,0.389981448713)(181,0.388782454484)(182,0.364307436457)(183,0.36167074885)(184,0.35933550029)(185,0.358806756263)(186,0.350392158857)(187,0.346636608053)(188,0.342509751404)(189,0.328479915974)(190,0.326128371986)(191,0.323823361525)(192,0.32182496478)(193,0.321389986744)(194,0.317732176288)(195,0.314134449238)(196,0.312601663558)(197,0.312195552015)(198,0.309319215534)(199,0.303839782836)(200,0.302835765942)(201,0.302081367789)(202,0.301790328023)(203,0.293061174621)(204,0.288142343366)(205,0.279091318581)(206,0.259090099354)(207,0.24241616004)(208,0.241589458635)(209,0.233399707931)(210,0.221343785042)(211,0.21582106454)(212,0.164650366753)(213,0.153289972964)};
\addlegendentry{StartPage}
\addplot coordinates {(1,0.869615544482)(2,0.868077384765)(3,0.864595986694)(4,0.863784975615)(5,0.863300089491)(6,0.860537035863)(7,0.856442464844)(8,0.856411307512)(9,0.856136821398)(10,0.855277141086)(11,0.855230097123)(12,0.85509775674)(13,0.854252050627)(14,0.853900559155)(15,0.852743652652)(16,0.852562494809)(17,0.851734939167)(18,0.850427017996)(19,0.850148520401)(20,0.847807409685)(21,0.846852485328)(22,0.846133598593)(23,0.845910826926)(24,0.844625648362)(25,0.844419668184)(26,0.84413381395)(27,0.843400941214)(28,0.84303233114)(29,0.842741177503)(30,0.842621613769)(31,0.842560468516)(32,0.841568380923)(33,0.840561285857)(34,0.839863899343)(35,0.83977345683)(36,0.83924030959)(37,0.83901783795)(38,0.836842825287)(39,0.83567945284)(40,0.835492060403)(41,0.835257313695)(42,0.833726366337)(43,0.833332587868)(44,0.832904402903)(45,0.832002613353)(46,0.831139755947)(47,0.830388590561)(48,0.829593475571)(49,0.829186599589)(50,0.829006411506)(51,0.827560228548)(52,0.826818580899)(53,0.826160567756)(54,0.825937658485)(55,0.825074283134)(56,0.823735387301)(57,0.820892035744)(58,0.818155889818)(59,0.81737606005)(60,0.815558644016)(61,0.813095707017)(62,0.811865470573)(63,0.811610237048)(64,0.811542503642)(65,0.811411652488)(66,0.810673172479)(67,0.810672592391)(68,0.810491856042)(69,0.810298226276)(70,0.809775546677)(71,0.809645951263)(72,0.808465706405)(73,0.808435318007)(74,0.807971852802)(75,0.807264051396)(76,0.807152823129)(77,0.806703455424)(78,0.805160653406)(79,0.8042953482)(80,0.80323881606)(81,0.803219981829)(82,0.802687946919)(83,0.802579363163)(84,0.801870742826)(85,0.801419966411)(86,0.800643206136)(87,0.79808161529)(88,0.798007109474)(89,0.797526742777)(90,0.797313825947)(91,0.79469843436)(92,0.79426280517)(93,0.793937410187)(94,0.79348040989)(95,0.791955883904)(96,0.790426205697)(97,0.789633578917)(98,0.788846248166)(99,0.788130936336)(100,0.787989682293)(101,0.787490646713)(102,0.784127113694)(103,0.782511019324)(104,0.777905600198)(105,0.775660456771)(106,0.775267896033)(107,0.774741459886)(108,0.774530889877)(109,0.774500437913)(110,0.770382871905)(111,0.770369671368)(112,0.767875915491)(113,0.767778033972)(114,0.767704416755)(115,0.767655925733)(116,0.76383184161)(117,0.761331543007)(118,0.754356510201)(119,0.750395454278)(120,0.740621807349)(121,0.738142374)(122,0.731812594769)(123,0.726834928012)(124,0.724329975026)(125,0.720777274711)(126,0.698507838451)(127,0.692986203149)(128,0.691468562795)(129,0.690114486404)(130,0.688961018594)(131,0.680788966605)(132,0.679384336646)(133,0.663959540206)(134,0.661500664146)(135,0.658228227689)(136,0.649907318512)(137,0.641694623521)(138,0.641263777534)(139,0.63511681716)(140,0.625075005101)(141,0.612850997254)(142,0.605392389963)(143,0.601448912261)(144,0.595075164128)(145,0.577945519796)(146,0.576013188384)(147,0.568718155042)(148,0.567492001049)(149,0.564931561344)(150,0.563840197605)(151,0.559918288444)(152,0.547506384096)(153,0.537150767344)(154,0.535512989905)(155,0.522252484197)(156,0.518984626215)(157,0.516961232675)(158,0.51585921119)(159,0.513048136401)(160,0.507654608572)(161,0.50743120244)(162,0.500499739841)(163,0.487910916193)(164,0.479873749038)(165,0.479611943639)(166,0.470452248919)(167,0.468627828638)(168,0.458908055251)(169,0.450769459658)(170,0.449278141687)(171,0.434180299993)(172,0.432773047985)(173,0.424133382442)(174,0.421753374745)(175,0.42054384107)(176,0.41452565966)(177,0.413875913989)(178,0.413001733121)(179,0.411759873403)(180,0.404490197611)(181,0.402881852591)(182,0.40102516013)(183,0.399853056376)(184,0.393743333952)(185,0.376118797282)(186,0.369875580109)(187,0.366321669418)(188,0.363367360116)(189,0.358907103402)(190,0.357921194452)(191,0.352483532834)(192,0.351340604198)(193,0.347095067804)(194,0.346455011856)(195,0.337719610564)(196,0.333094794331)(197,0.328484535733)(198,0.306236710715)(199,0.301838516949)(200,0.300088869867)(201,0.285501799615)(202,0.274684303597)(203,0.266028109168)(204,0.26407321685)(205,0.231808958638)(206,0.206287103366)(207,0.195746567933)(208,0.184440110501)(209,0.180062503565)(210,0.171203067266)(211,0.156441813383)(212,0.153319738035)(213,0.14685129993)};
\addlegendentry{Qwant}
\addplot coordinates {(1,1.0)(2,1.0)(3,1.0)(4,1.0)(5,1.0)(6,1.0)(7,1.0)(8,1.0)(9,1.0)(10,1.0)(11,1.0)(12,1.0)(13,1.0)(14,1.0)(15,1.0)(16,1.0)(17,1.0)(18,1.0)(19,1.0)(20,1.0)(21,1.0)(22,1.0)(23,1.0)(24,1.0)(25,1.0)(26,1.0)(27,1.0)(28,1.0)(29,1.0)(30,1.0)(31,1.0)(32,1.0)(33,1.0)(34,1.0)(35,1.0)(36,1.0)(37,1.0)(38,1.0)(39,1.0)(40,1.0)(41,1.0)(42,1.0)(43,1.0)(44,1.0)(45,1.0)(46,1.0)(47,1.0)(48,1.0)(49,1.0)(50,1.0)(51,1.0)(52,1.0)(53,1.0)(54,1.0)(55,1.0)(56,1.0)(57,1.0)(58,1.0)(59,1.0)(60,1.0)(61,1.0)(62,1.0)(63,1.0)(64,1.0)(65,1.0)(66,1.0)(67,1.0)(68,1.0)(69,1.0)(70,1.0)(71,1.0)(72,1.0)(73,1.0)(74,1.0)(75,1.0)(76,1.0)(77,1.0)(78,1.0)(79,1.0)(80,1.0)(81,1.0)(82,1.0)(83,1.0)(84,1.0)(85,1.0)(86,1.0)(87,1.0)(88,1.0)(89,1.0)(90,1.0)(91,1.0)(92,1.0)(93,1.0)(94,1.0)(95,1.0)(96,1.0)(97,1.0)(98,1.0)(99,1.0)(100,1.0)(101,1.0)(102,1.0)(103,1.0)(104,1.0)(105,1.0)(106,1.0)(107,1.0)(108,1.0)(109,1.0)(110,1.0)(111,1.0)(112,1.0)(113,1.0)(114,1.0)(115,1.0)(116,1.0)(117,1.0)(118,1.0)(119,1.0)(120,1.0)(121,1.0)(122,1.0)(123,1.0)(124,1.0)(125,1.0)(126,1.0)(127,1.0)(128,1.0)(129,1.0)(130,1.0)(131,1.0)(132,1.0)(133,1.0)(134,1.0)(135,1.0)(136,1.0)(137,1.0)(138,1.0)(139,1.0)(140,1.0)(141,1.0)(142,1.0)(143,1.0)(144,1.0)(145,1.0)(146,1.0)(147,1.0)(148,1.0)(149,1.0)(150,1.0)(151,1.0)(152,1.0)(153,1.0)(154,1.0)(155,1.0)(156,1.0)(157,1.0)(158,1.0)(159,1.0)(160,1.0)(161,1.0)(162,1.0)(163,1.0)(164,1.0)(165,1.0)(166,1.0)(167,1.0)(168,1.0)(169,1.0)(170,1.0)(171,1.0)(172,1.0)(173,1.0)(174,1.0)(175,1.0)(176,1.0)(177,1.0)(178,1.0)(179,1.0)(180,1.0)(181,1.0)(182,1.0)(183,1.0)(184,1.0)(185,1.0)(186,1.0)(187,1.0)(188,1.0)(189,1.0)(190,1.0)(191,1.0)(192,1.0)(193,1.0)(194,1.0)(195,1.0)(196,1.0)(197,1.0)(198,1.0)(199,1.0)(200,1.0)(201,1.0)(202,1.0)(203,1.0)(204,1.0)(205,1.0)(206,1.0)(207,1.0)(208,1.0)(209,1.0)(210,1.0)(211,1.0)(212,1.0)(213,1.0)};
\addlegendentry{Consensus}
\end{axis} 
\end{tikzpicture}
\caption{Distribution of scores relative to the consensus, from largest to smallest.}
\label{fig:distribrelativescore}
\end{figure}
\end{center}
It allows to see the variations of score per SE. Again the same SE groups appear, but the information is stronger than just the mean. For the first quarter of requests, scores are close for all SEs except Ask.com, the difference becomes significant later with some  SEs which cannot keep up with he best ones.

To identify deviations from other search engines, we highlight respectively in Tables~\ref{tab:high} and~\ref{tab:low} for each SE the (ordered) 10 queries with the highest and lowest relative score with respect to the consensus SE.
Those queries correspond to the extreme left (for \cref{tab:high}) and extreme right (for \cref{tab:low}) of \cref{fig:distribrelativescore}.
\begin{table}[bht]
\begin{center}\footnotesize
\begin{tabular}{p{.1\textwidth}p{.1\textwidth}p{.1\textwidth}p{.1\textwidth}p{.1\textwidth}p{.1\textwidth}p{.1\textwidth}p{.1\textwidth}p{.1\textwidth}}
Google & Yahoo & Bing & AOL & Ask & DuckDuckGo & Ecosia & StartPage & Qwant\\
\hline\hline
(0.8667) & (0.8799) & (0.8835) & (0.8799) & (0.4915) & (0.8821) & (0.8821) & (0.8667) & (0.8696) \\
how many days until christmas & how to cook quinoa & how to cook quinoa & how to cook quinoa & what does hmu mean & how to take a screenshot on a mac & how to take a screenshot on a mac & how many days until christmas & cricbuzz \\\hline
(0.8635) & (0.8793) & (0.8815) & (0.8793) & (0.3301) & (0.8813) & (0.8818) & (0.8635) & (0.8681) \\
how much house can i afford & how to take a screenshot on a mac & how much house can i afford & how to take a screenshot on a mac & how to draw a doghow to get rid of blackheads & how many days until christmas & MercadoLibre & how much house can i afford & how much house can i afford \\\hline
(0.8626) & (0.8792) & (0.8742) & (0.8792) & (0.2802) & (0.8799) & (0.8813) & (0.8568) & (0.8646) \\
how many days till christmas & what time is sunset & how to take a screenshot on a mac & what time is sunset & craigslist & how to cook quinoa & how many days until christmas & how many days till christmas & ebay kleinanzeigen \\\hline
(0.8581) & (0.8782) & (0.8728) & (0.879) & (0.2756) & (0.8796) & (0.8806) & (0.8557) & (0.8638) \\
omegle & speedometer test & how to take a screenshot & MercadoLibre & who sings this song & crikbuzz & how to screenshot on mac & what is my ip address & how to write a cover letter \\\hline
(0.8559) & (0.8782) & (0.8724) & (0.8788) & (0.2694) & (0.8792) & (0.8799) & (0.8513) & (0.8633) \\
what is my ip address & what is my ip & how to download videos from youtube & crikbuzz & how to make french toast & what time is sunset & flipkart & home-depot & crikbuzz \\\hline
(0.8531) & (0.878) & (0.8713) & (0.8783) & (0.2649) & (0.8782) & (0.879) & (0.85) & (0.8605) \\
national basketball association & cricbuzz & how many centimeters in an inch & what is my ip & restaurant & flipkart & what time is sunset & what is my ip & how many ounces in a liter \\\hline
(0.852) & (0.8779) & (0.8688) & (0.8766) & (0.2509) & (0.8782) & (0.8788) & (0.8479) & (0.8564) \\
when we were young & national basketball association & what time is it in california & how many mb in a gb & tiempos & irctc & weather & national basketball association & how to take a screenshot on a mac \\\hline
(0.85) & (0.8773) & (0.8681) & (0.8762) & (0.2442) & (0.8782) & (0.8774) & (0.8466) & (0.8564) \\
what is my ip & weather & what time is it in london & how to write a check & amazon & what is my ip & how many days till christmas & how old is justin bieber & juegos \\\hline
(0.8479) & (0.8771) & (0.867) & (0.8751) & (0.2421) & (0.8774) & (0.8774) & (0.8461) & (0.8561) \\
euro 2016 & what time is it in london & why is the sky blue & what is my ip address & mailen & how much house can i afford & how to cook quinoa & how to take a screenshot on a mac & what time is it in london \\\hline
(0.8466) & (0.8768) & (0.8668) & (0.8745) & (0.2402) & (0.8772) & (0.877) & (0.8449) & (0.8553) \\
how many people are in the world & why is the sky blue & crikbuzz & weather & national basketball association & speedometer test & tubemate & how many people are in the world & bed 365 \\\hline

\end{tabular}
\caption{Per SE, ordered list of 10 queries with the largest relative score with respect to the consensus (and their relative scores)}\label{tab:high} 
\end{center}
\end{table}
\begin{table}[bht]
\begin{center}\footnotesize
\begin{tabular}{p{.1\textwidth}p{.1\textwidth}p{.1\textwidth}p{.1\textwidth}p{.1\textwidth}p{.1\textwidth}p{.1\textwidth}p{.1\textwidth}p{.1\textwidth}}
Google & Yahoo & Bing & AOL & Ask & DuckDuckGo & Ecosia & StartPage & Qwant\\
\hline\hline
(0.2219) & (0.2686) & (0.1713) & (0.2576) & (0.115) & (0.4272) & (0.2803) & (0.1533) & (0.1469) \\
convertidos & yahoomail & skype & when is fathers day & how to take a screenshot on a mac & traduttor & bbc news & beeg & google maps \\\hline
(0.2309) & (0.44) & (0.2025) & (0.2581) & (0.1153) & (0.5306) & (0.3822) & (0.1647) & (0.1533) \\
how to draw a doghow to get rid of blackheads & mail & ikea & yahoomail & what is my ip address & how many ounces in a quart & where are you now & convertidos & minecraft \\\hline
(0.2322) & (0.5195) & (0.2287) & (0.2609) & (0.1179) & (0.5929) & (0.4572) & (0.2158) & (0.1564) \\
restaurant & how to make love & gmail & traductor google & how to screenshot on mac & how to start a business & how to make money & daily mail & msn \\\hline
(0.2564) & (0.5607) & (0.2301) & (0.2862) & (0.1183) & (0.6261) & (0.4779) & (0.2213) & (0.1712) \\
how many ounces in a quart & who sings this song & youtube & when is mothers day & football association & messenger & games & omegle & news \\\hline
(0.2686) & (0.6264) & (0.2318) & (0.2936) & (0.1188) & (0.6319) & (0.5005) & (0.2334) & (0.1801) \\
cnn & messenger & youtube mp3 & where are you now & juegos & hotmail & how tall is kevin hart & restaurant & google drive \\\hline
(0.28) & (0.6351) & (0.2332) & (0.3147) & (0.119) & (0.6739) & (0.5194) & (0.2416) & (0.1844) \\
ryanair & oranges & pokemon go & mail & how many centimeters in an inch & who sings this song & tiempos & how to make love & outlook \\\hline
(0.2957) & (0.6552) & (0.2452) & (0.3156) & (0.1192) & (0.6757) & (0.5219) & (0.2424) & (0.1957) \\
putlocker & how do you spell & ryanair & messenger & irctc & euro 2016 & myn & ryanair & skype \\\hline
(0.3033) & (0.6756) & (0.2928) & (0.3196) & (0.1193) & (0.7268) & (0.5238) & (0.2591) & (0.2063) \\
what time is it in australia & euro 2016 & aleg & what is your name & how much house can i afford & zalando & how to make money fast & mincraft & zara \\\hline
(0.3038) & (0.6794) & (0.2981) & (0.427) & (0.1193) & (0.7293) & (0.528) & (0.2791) & (0.2318) \\
instagram & what is the temperature & facebook & euro 2016 & weather & what is the temperature & how old is hillary clinton & cnn & youtube mp3 \\\hline
(0.3059) & (0.7037) & (0.2981) & (0.4532) & (0.1196) & (0.7296) & (0.5438) & (0.2881) & (0.2641) \\
traduttore & how many weeks in a year & pandora & who sings this song & how many people are in the world & how to make pancakes & when we were young & how to draw a doghow to get rid of blackheads & gmail \\\hline
\end{tabular}
\caption{Per SE, ordered list of 10 queries with the smallest relative score with respect to the consensus (and their relative score).}\label{tab:low} 
\end{center}
\end{table}

Search terms displayed in \cref{tab:high} appear to be quite complex--or specific--searches, for which there is not much room for disagreement among SEs.

On the other hand, \cref{tab:low} shows, for each SE, the terms for which they most disagree with the consensus, which may help highlight non-neutral behaviors. For example, it is interesting to note that Bing, the Microsoft-operated SE, differs most from the consensus (hence, from the other SEs) on some sensitive searches such as \word{skype, gmail, youtube}, and \word{facebook}. Similarly, AOL strongly disagrees with the consensus for \word{yahoomail, mail}, and \word{messenger}. While Qwant gets low scores for \word{google maps, msn, outlook, news, google drive, skype}, and \word{gmail}, all involving SE-controlled services. Finally, let us note that we may also have a look at searches like \word{restaurant} or \word{cnn}, for which Google is far from the consensus: is that to favor its own news and table-booking services? Our study is too preliminary to draw definite conclusions, but can help raise such questions.


\section{Conclusions} \label{sec:conc}

In this paper, we have defined a measure of relevance of web pages for given queries based on the visibility/response from a whole set of search engines. This relevance takes into account the position of the web page thanks to a weight corresponding to the click-through-rate of the position. It then allowed to define a score of a search engine for a given query, and the average score for a whole set of queries. 

We designed a tool in Python allowing to study the scores of nine known search engines and to build the consensus ranking maximizing the SE score, for a set of more that two hundred queries. 
A first analysis suggests that there are significant differences among search engines, that may help identify some sensitive terms subject to biases in the rankings.
 
We finally note that our method does not provide an absolute-value score for each SE allowing to decide which is the best one, but rather indicates whether an SE agrees with the others. The user may very well prefer an SE that is far from our consensus ranking, especially if that SE better takes her preferences into account when performing the ranking.

In a follow up of this preliminary work, we plan to design statistical tests of potentially intentional deviations by search engines from their regular behavior, to highlight if non-neutrality by search engines can be detected and harm some content providers. This would be a useful tool within the search neutrality debate.


\end{document}